\documentstyle[aps,preprint,psfig]{revtex}
\begin{document}
\draft
\preprint{Submitted for publication in the ECOSS Proceedings, 9-13 Sept. 1996, Genova, Surface Science}
\title{Study of CO Oxidation over Ru\,(0001) at High Gas Pressures}
\author{C. Stampfl and M. Scheffler}
\address{
Fritz-Haber-Institut der Max-Planck-Gesellschaft,
Faradayweg 4-6, D-14 195 Berlin-Dahlem, Germany
}
\date{\today}


\maketitle

\vspace*{-10pt}
\vspace*{-0.7cm}
\begin{quote}
\parbox{16cm}{\small
Experiments performed at high gas partial pressures
have demonstrated that the kinetics of the CO oxidation reaction
at Ru\,(0001) is different and somewhat anomalous compared
to that over other transition metal
surfaces and, in particular, 
the turnover rate is exceptionally high.
In order to gain insight into the underlying reasons for this behavior,
we performed density functional theory calculations using the 
generalized gradient approximation for the exchange-correlation functional.
We find that the high rate is due
to a weakly, but nevertheless well bound, $(1 \times 1)$ oxygen adsorbate 
layer which may form for high O$_{2}$ pressures  but not under
usual ultra high vacuum conditions. The calculations indicate that
reaction to CO$_{2}$ occurs both via scattering of 
{\em gas-phase} CO molecules as well as by
CO molecules weakly  adsorbed
at vacancies in the oxygen adlayer, where the latter mechanism
dominates the rate.\\
} 
\end{quote}
 
\section{Introduction}
The oxidation of carbon monoxide at transition metal surfaces
is one of the most extensively studied and best understood
heterogeneous catalytic reactions (see, for example
Refs.~\cite{king} and references therein).
Nevertheless, very little in fact, is actually known about the reaction pathway
 on a {\em microscopic} level.
Recently there has been a resurge of interest in such surface chemical
reactions which
has been stimulated by advances in surface science techniques (e.g.
molecular and atomic beam experiments~\cite{rendulic}
and high pressure catalytic reactors \cite{goodman}).
These techniques have 
enabled new information to be obtained concerning the behavior
of chemisorbed reaction partners.
Recent studies using high gas pressure catalytic reactors
(e.g. operating  at $\sim$ 10 torr with CO/O$_{2}$ pressure ratios $ < 1$)
have reported unusual behavior for the CO oxidation reaction
over Ru\,(0001) \cite{peden1}: 
$(i)$ The rate of CO$_{2}$ production
is significantly higher than at other
transition metal surfaces; in contrast, under ultra high 
vacuum (UHV) conditions, Ru\,(0001) is notably
the poorest catalyst for this reaction~\cite{king}. 
$(ii)$ The measured kinetic data (e.g., activation energy, and
temperature and pressure dependencies of the rate)
is markedly different to that of
other substrates. $(iii)$ As opposed to other transition metal 
catalysts (Pt, Pd, Ir, and Rh), 
 highest rates of CO$_{2}$ formation
 occur for high surface oxygen concentrations (one monolayer was
proposed on the basis of auger electron spectroscopy measurements).
 $(iv)$ Almost no chemisorbed CO was detected during or after
the reaction.  On the basis of these findings,
it was speculated that an Eley-Rideal (E-R) mechanism is operational 
as opposed to the ``usual'' Langmuir-Hinshelwood (L-H) mechanism.
In the E-R mechanism, the reaction occurs  between
{\em gas-phase} and chemisorbed particles, while in the L-H mechanism,
the reaction occurs between species both chemisorbed on and in 
thermal equilibrium with the surface.
So far E-R mechanisms have only recently been experimentally
confirmed, and in most cases they are for
somewhat artificial reactions between gas-phase
atomic hydrogen or deuterium (from a beam source) with chemisorbed 
atoms~\cite{rettner}. 

In the present paper we describe results our theoretical study
aimed at understanding the experimental findings
described above for the oxidation of CO at Ru\,(0001).

\section{Calculation Method}
We performed density functional theory (DFT)
 calculations using 
the pseudopotential plane wave method and the supercell approach
\cite{stumpf}.
Surfaces were modelled by  unit cells of $(2 \times 2)$
 and $(1 \times 1)$ periodicity
with four layers of Ru\,(0001) and a vacuum region equivalent
to thirteen such layers.
The {\em ab initio} fully separable pseudopotentials were
created using the scheme of Troullier and
Martins~\cite{troullier}. Plane wave cut-offs of 40 Ry, 
with three {\bf k}-points \cite{cunningham} in the surface Brillouin zone 
of the $(2 \times 2)$ cell were used, as well as a 60 Ry cut-off with
fifty-two {\bf k}-points in the surface Brillouin 
zone of the $(1 \times 1)$ cell.
Results of calculations performed using these different basis sets indicated
that the smaller basis set provides a sufficiently accurate description
for the present investigation~\cite{stampfl1}.
The generalized gradient approximation (GGA) of Perdew 
{\em et al.}~\cite{perdew} was employed
for the exchange-correlation functional and it was used in creating
the pseudopotential
as well as in the total-energy functional. It is therefore
treated in a consistent way.
The position of all atoms were allowed to relax except the bottom two Ru 
layers which were held at their bulk positions.
We like to point out that the present study represents 
the first investigation of a
{\em surface chemical reaction} (to be distinguished from dissociative 
chemisorption)
 using such theoretical methods.
Further details about the calculations can
be found in Refs.~\cite{stampfl1,stampfl2,over}.

\section{Adsorption of O on Ru\,(0001)}
Given that the CO oxidation experiments identified that highest
rates of CO$_{2}$ production occurred for high O$_{2}$ partial pressures and  
concomitantly high oxygen coverages,
an important part of the theoretical study is therefore
to investigate the structure and stability of high coverage
O adlayers on Ru\,(0001).
It is known that using O$_{2}$, under UHV conditions at 300~K,
the saturation coverage is
close to $\Theta=0.5$.
Earlier we reported the results of  
DFT-GGA calculations for various adsorbate
phases of oxygen on Ru\,(0001)
and found that an even higher coverage should be 
stable on the surface~\cite{stampfl1}. In particular, a 
structure with a ($1 \times 1$) periodicity
and coverage  $\Theta = 1$ where 
the oxygen atoms occupy the hcp-hollow sites. 
(The coverage $\Theta$
is defined as the ratio of the
concentration of adparticles to that of substrate atoms in the topmost
layer.)
This result indicated that
the formation of the $\Theta = 1$ structure under UHV conditions
from gas phase O$_2$ is kinetically hindered, 
but that by offering atomic oxygen, this phase should be attainable.
Indeed this structure was successfully created
by starting from the $\Theta=1/2$ phase
and offering additional
{\em atomic} oxygen to the surface via the dissociation of NO$_2$~\cite{over}.
The determined surface atomic structure of the $(1 \times 1)$ phase by 
the low-energy electron diffraction (LEED) intensity analysis
of Ref.~\cite{over} 
agreed very well with that predicted by the DFT-GGA calculations.
The calculations show that the adsorption
energy of oxygen in this phase is significantly weaker than it is in the
lower coverage ordered phases (of $(2 \times 2)$ and $(2 \times 1)$
periodicities).
This can be seen from Fig.~1 which 
shows the adsorption energy, with respect to O$_{2}$ gas,
for oxygen in the hcp- and fcc-hollow sites at various coverages.
The region over which dissociative chemisorption of oxygen is
kinetically hindered is indicated as the shaded region.
It can be seen that the hcp-hollow site is favored at all coverages, where
the adsorption energy of the $(2 \times 2)$ phase is the greatest. This 
suggests that at low coverages island formation will occur 
with a $(2 \times 2)$ periodicity which is consistent with experimental
observations~\cite{kostov,wintterlin}. With subsequently
increasing coverage, the adsorption energy decreases which reflects
a repulsive adsorbate-adsorbate interaction, and indicates that
no island formation will take place in the coverage regime $\Theta=0.25$ to 1.
Furthermore, with increasing coverage from $\Theta=0.25$,
the difference in adsorption
energy between the fcc- and hcp-hollow sites decreases, and at coverage one
the difference is very small ($\sim$0.06~eV).

Catalysis involving high O$_{2}$ pressures is therefore likely
to involve oxygen coverages that approach one monolayer on the surface
because the impingement rate is proportional to the partial pressure
which means that there is a significantly higher attempt frequency 
to overcome activation barriers for dissociative adsorption
than under UHV conditions.
We therefore initially assumed in our investigation of the 
oxidation of CO at Ru\,(0001) 
that the $(1 \times1 )$ phase covers the surface.

\section{Oxidation of CO at $(1 \times 1)$-O/Ru\,(0001)}

For the catalytic oxidation of CO, it is 
necessary that a CO molecule reacts with
an oxygen atom to yield CO$_{2}$.
 From the experiments it had been speculated that  CO 
may achieve this from the gas-phase without adsorbing on the surface first, i.e.
the E-R mechanism. To investigate the possibility of this scenario, calculations
were  first carried out to determine whether CO could adsorb on the 
$(1 \times 1)$ oxygen covered Ru\,(0001) surface.
The sites considered were the on-top and  fcc-hollow sites,
with respect to the Ru\,(0001) substrate, and a bridge site between two adsorbed
O atoms. It was found
that at all these sites, CO is unstable, thus {\em preventing} the L-H mechanism
on the perfectly ordered $(1 \times 1)$-O/Ru\,(0001) surface.

To investigate the possibility of reaction via an E-R mechanism,
we evaluate an appropriate cut through the high-dimensional
potential energy surface (PES); 
this cut is defined by two variables: the
vertical position of the C atom 
and the vertical position of the O adatom below the molecule.
In order of ease of analysis,
the CO axis is held perpendicular to the surface.
The resulting PES  is presented in Fig.~2 where
the coordinate system is shown as the inset. 
The most favorable pathway can quickly be seen: it
has an energy barrier of about 1.6~eV and formation of 
CO$_{2}$ is achieved via an upward movement
of the O adatom by $\approx 0.4$ \AA\, towards the CO molecule. 
Due to the similar masses of O and C, it is likely that the
impinging CO molecule will impart a significant amount of energy to the O
adatom, thus stimulating its vibrations and facilitating its motion
(indicated by the oscillations in the dot-dashed line).
This would then  bring the system to the transition state of the reaction
(marked by the asterisk).
The newly formed CO$_{2}$ molecule is strongly repelled
from the surface and moves into the gas phase region 
with a large energy gain of 1.95~eV.

As mentioned, the PES of Fig.~2 corresponds to a constrained
situation of the surface--CO angle. When this constraint is dropped, i.e.,
when the tilt angle of the CO axis
is allowed to relax, we find that
the energy barrier is reduced to 1.1~eV, and
also that the position of the saddle point occurs closer  to the
surface (by 0.3~\AA).  
The geometry of the identified  transition state at the
saddle point is depicted in Fig.~3. 
The optimum tilt angle with respect to the
surface normal is  $49^{\rm o}$ which 
corresponds to a ``bond angle''
of  $131^{\circ}$ for the ``CO$_{2}$-like'' complex.
It is interesting to note that this geometry is similar to that associated
with the CO$_{2}^{-}$ ion \cite{bagus} and to that proposed
for the ``activated complex'' for the CO oxidation reaction over
other transition metal surfaces \cite{coulston} in which the reaction
proceeds via the L-H mechanism.

In Fig.~4 the energy diagram for the proposed 
reaction pathway for the E-R mechanism is given.
Interestingly,
the calculations show that there is a physisorption
well  for CO (over all of the surface unit cell),
as well as for CO$_{2}$, above the surface.
The calculated depths however ($\sim$0.04~eV), are likely to be underestimated
because the employed
exchange-correlation functional does not describe the long-range
van der Waals type interactions.
It can be seen that with respect to the free CO and $\frac{1}{2}$O$_{2}$
there is an energy gain of 1.20~eV on adsorption of oxygen
into a vacancy in the $(1 \times 1)$ structure, 
an energy barrier for reaction to CO$_{2}$, and a significant energy
gain on formation of CO$_{2}$ which leaves  an oxygen vacancy
 behind on the surface.
This predicted  mechanism and associated energetics could possibly be tested by 
molecular beam experiments. It is, however, unlikely that
this process alone can explain the high rate measured in the high pressure
catalytic reactor experiments  given the
height of the activation energy barrier, which is
high for an E-R mechanism.

It is conceivable that 
CO molecules adsorb at vacancies in the O-adlayer, i.e., at sites 
at which
an oxygen atom has been removed (e.g. by the above described E-R reaction).
We find that CO can weakly adsorb in such a vacancy where the adsorption
energy is 0.85~eV-- significantly less than on the clean
surface which we calculate to be 1.68~eV 
(the experimental value has been determined to be 1.66~eV~\cite{coexp}).
Adsorption of $\frac{1}{2}$O$_{2}$ in such a vacancy is more favorable, with
an adsorption energy of 1.20~eV. 
Using these adsorption energies of CO and $\frac{1}{2}$O$_{2}$, and
assuming thermal equilibrium of the
CO + O$_2$ gas and a mixed CO + O adlayer, the law of mass
action indicates 
that about 0.03~\% of the sites of the $(1 \times 1)$ adlayer will be
occupied by CO. (We assumed that the
O$_2$ and CO partial pressures are equal and the temperature
is $T = 500$~K.) 
In reality the CO concentration will be even higher
because we find that CO adsorption into an existing O vacancy
can proceed basically without hindrance while O adsorption (from O$_2$)
is hindered by an energy barrier.
Therefore the actual percentage of surface sites occupied by CO will 
be somewhat higher.
A L-H reaction between the adsorbed CO molecule in a vacancy
and a neighboring O adatom is expected to be
particularly efficient and to give rise to the high rate
 due to the weaker bond strengths of
both the adsorbed CO molecule and O adatom, as well as their  close proximity 
to one another.
In contrast, under UHV conditions, where the
high coverage phase is unable to form due to 
kinetic hindrance for O$_{2}$ dissociation, Ru\,(0001) is the 
poorest catalyst for this reaction. The reason for this is likely to be 
related to the fact that under these conditions
 ruthenium binds oxygen
(and carbon monoxide) particularly strongly.


\newpage

\begin{figure}
\psfig{figure=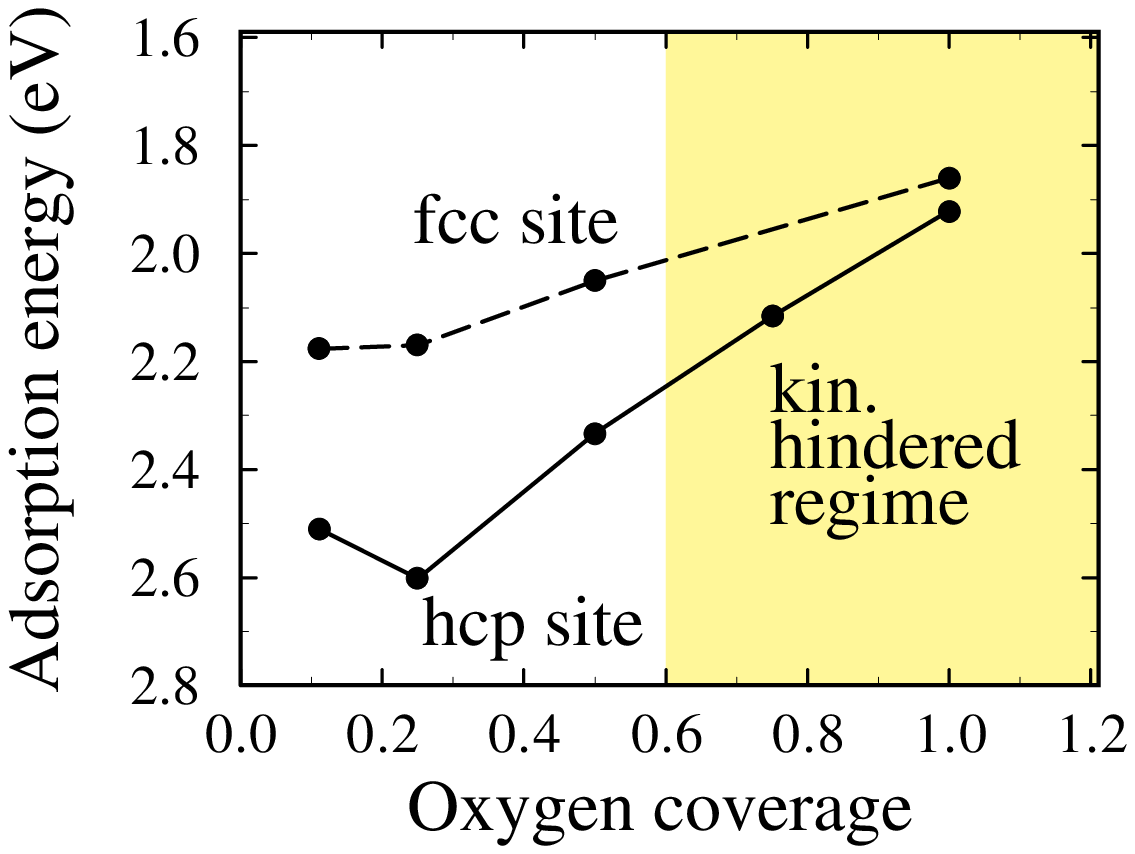}
\caption{
Adsorption energy of O on Ru\,(0001) with respect to $\frac{1}{2}$O$_2$
at various coverages for oxygen in the fcc- (dashed line)
and hcp-hollow sites (continuous line).  }
\end{figure}

\clearpage

\begin{figure}
\psfig{figure=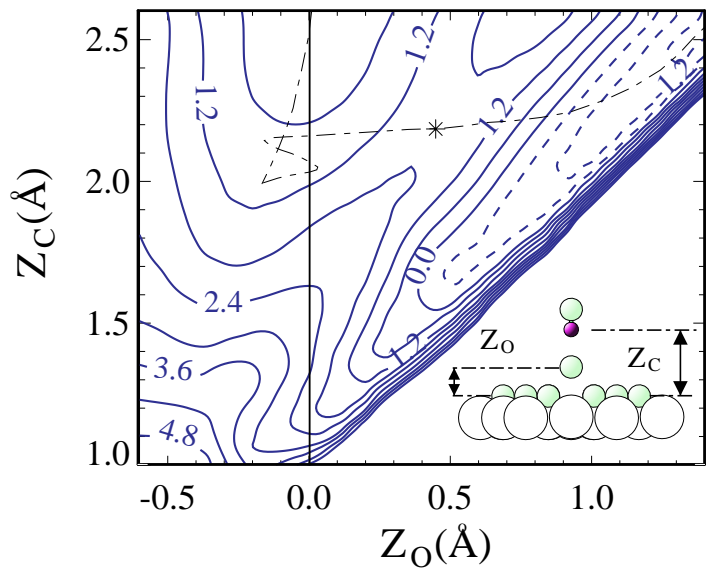}
\caption{Potential energy surface (PES) 
as a function of the positions of the C atom, $Z_{\rm C}$, 
and of the O adatom, $Z_{\rm O}$.
Positive energies are shown as dashed lines, negative ones as
full lines.
The contour line spacing is 0.6 eV.  }
\end{figure}
\clearpage

\begin{figure}
\psfig{figure=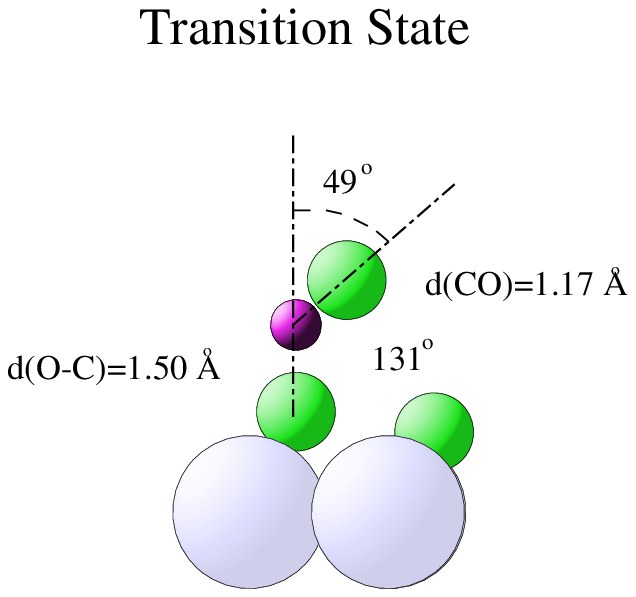}
\caption{Illustration of the transition state identified 
for the  reaction of gas-phase CO
with adsorbed oxygen when tilting of the CO axis is considered. }
\end{figure}

\begin{figure}
\psfig{figure=fig4.ps}
\caption{
Predicted energy diagram for the E-R mechanism of CO oxidation
at Ru\,(0001). Note that the depths of the physisorption wells are
exaggerated for visibility. }
\end{figure}

\end{document}